\documentclass[conference]{IEEEtran}
\usepackage[letterpaper, left=1in, right=1.05in, bottom=1in, top=0.75in]{geometry}

\usepackage{amsfonts}
\usepackage{rotating}
\usepackage{etex}
\usepackage{amsmath,amssymb}
\usepackage{mathtools}
\usepackage{amsthm}
\usepackage{float}
\usepackage{slashbox}
\usepackage{subcaption}
\usepackage{nicefrac}
\usepackage[dvipsnames]{xcolor}
\usepackage{tikz}
\usepackage{ctable}
\usepackage{graphicx}  
\usepackage{tabularx}
\usepackage{accents}
\usepackage{enumitem}
\usepackage[makeroom]{cancel}
\usepackage{colortbl}
\usepackage{multirow,balance}
\usepackage{mathtools}
\usepackage{pgfplots}
\pgfplotsset{compat=newest}
\usetikzlibrary{plotmarks}
\usetikzlibrary{spy}
\usepackage{grffile}
\usepackage{amsmath}

\usepackage{balance}
 
\allowdisplaybreaks % used for multi-page setup of equations 

\usetikzlibrary{decorations.text}
\usepgfplotslibrary{fillbetween}
\usetikzlibrary{calc, fit}
\usetikzlibrary{arrows.meta}
\usetikzlibrary{patterns}
\usetikzlibrary{fit,matrix}

\newcommand{\iPhone}[3]{
	\coordinate (a) at (#1,#2);
	\draw [line width=0.25pt,rounded corners=(#3)*1mm,fill=white,scale=(#3)] (a)--($(a)+(0.67,0)$)--($(a)+(0.67,1.381)$)--($(a)+(0,1.381)$)--cycle;
	\draw [color=gray,line width=0.25pt,rounded corners=(#3)*0.8mm,fill=white,scale=(#3)]($(a)+(0.015,0.015)$)--($(a)+(0.655,0.015)$)--($(a)+(0.655,1.366)$)--($(a)+(0.015,1.366)$)--cycle;
	\draw [line width=0.25pt,rounded corners=(#3)*0.04mm,scale=(#3)] ($(a)+(0.2875,1.266)$)--($(a)+(0.3825,1.266)$)--($(a)+(0.3825,1.281)$)--($(a)+(0.2875,1.281)$)--cycle;
	\draw[line width=0.25pt,scale=#3] ($(a)+(0.335,0.09)$) circle (0.055cm);
	\draw[line width=0.25pt,scale=#3] ($(a)+(0.335,0.09)$) circle (0.044cm);
	\draw[line width=0.25pt,scale=#3] ($(a)+(0.2275,1.2735)$) circle (0.015cm);
	\draw[line width=0.25pt,scale=#3] ($(a)+(0.335,1.32)$) circle (0.01cm);
	\draw [fill={rgb:black,1;white,4},line width=0.25pt,scale=(#3)] ($(a)+(0.042475,0.170195)$)--($(a)+(0.042475,0.170195)+(0.58505,0.0)$)--($(a)+(0.042475,0.170195)+(0.58505,1.04061)$)--($(a)+(0.042475,0.170195)+(0.0,1.04061)$)--cycle;
}

\newcommand{\SymModNew}[0]{
	\begin{tikzpicture}[
	roundnode/.style={circle, draw=green!60, fill=green!5, very thick, minimum size=7mm},
	squarednode/.style={rectangle, draw=red!60, fill=red!5, very thick, minimum size=5mm},
	scale=1.25]
	
	\iPhone{0.5}{0}{0.5};
	\iPhone{3.1}{0}{0.5};
	\node (uav) at (1.8, 3.2) {\includegraphics[scale=0.35]{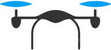}};
	
	\node[rectangle, fill=white, scale=0.8, align=center] (n1) at (3.2,3.2) [inner sep=0, align=left] {Intermittent \\ Node 1 (UAV)};	
	\node[rectangle, fill=white, scale=0.8, align=center] (n2) at (0,0.3) [inner sep=0] {Node 2\\(D2D)};	
	\node[rectangle, fill=white, scale=0.8, align=center] (n3) at (3.95,0.3) [inner sep=0] {Node 3\\(D2D)};
	
	\draw [->] (0.9, 0.45) to [out=20,in=160] (3.0,0.45);
	\draw [->] (3.0, 0.25) to [out=-160,in=-20] (0.9,0.25);
	\draw [->] (0.55, 0.75) to (1.675,2.45);
	\draw [->] (1.925,2.45) to (0.8, 0.75);
	\draw [->] (3.15, 0.75) to (1.675,2.45);
	\draw [->] (1.925,2.45) to (3.4, 0.75);
	
	\node[rectangle, fill=white, scale=0.8] (h31) at (2.5,1.4) [inner sep=0, scale=0.95] {$\boldsymbol{H}_{31}$};
	\node[rectangle, fill=white, scale=0.8] (h13) at (3,1.4) [inner sep=0, scale=0.95] {$\boldsymbol{H}_{13}$};	
	\node[rectangle, fill=white, scale=0.8] (h21) at (0.825,1.4) [inner sep=0, scale=0.95] {$\boldsymbol{H}_{21}$};
	\node[rectangle, fill=white, scale=0.8] (h12) at (1.325,1.4) [inner sep=0] {$\boldsymbol{H}_{12}$};
	\node[rectangle, fill=white, scale=0.8] (h23) at (1.895,0.625) [inner sep=0] {$\boldsymbol{H}_{23}$};
	\node[rectangle, fill=white, scale=0.8] (h32) at (1.895,0.05) [inner sep=0] {$\boldsymbol{H}_{32}$};			
	
	\node[fill=black, shape=circle, scale=0.4] (d1) at (1.675,2.5) {};
	\node[fill=black, shape=circle, scale=0.4] (d2) at (1.925,2.5) {};
	\draw[-] (1.675, 2.8+0.1) -- (1.675, 2.725+0.1) -- (1.53, 2.45+0.1);
	\draw[-] (1.895, 2.8+0.1) -- (1.895, 2.725+0.1) -- (1.75, 2.45+0.1);
	\draw[-, dashed] (1.25, 2.7) -- (2.25, 2.7);
	\draw[-, dotted] (1.55, 2.56) -- (2, 2.56);
	\node[rectangle, fill=white, scale=0.8] (sl) at (1.25,2.7) [inner sep=0] {$s_\ell$};		
	
	\end{tikzpicture}
}

\newcommand{\PlotLoS}[0]{
	\begin{axis}[
		axis x line=bottom, axis y line=left,
		ymin=0, ymax=1.075, %ytick={0.1,0.5, 0.9, 1}, 
		ytick={1},ylabel={$\mathbb{P}(\text{LoS})$},
		xmin=0, xmax=90, % xtick={37.5}, 
		xlabel={Elevation Angle $\theta\:[^{\circ}]$}, %grid
		]
		
		\addplot[name path=f1, very thick, black, domain=0:90, smooth]{1/(1+5.0188*exp(-0.3511*(x-5.0188)))} node[above, pos=0.35] {Suburban};
		
		\addplot[name path=f2, very thick, blue, domain=0:90, smooth]{1/(1+9.6101*exp(-0.1592*(x-9.6101)))} node[above, rotate=73, pos=0.25] {Urban};

		\addplot[name path=f3, very thick, cyan, domain=0:90, smooth]{1/(1+11.9480*exp(-0.1359*(x-11.9480)))} node[below, rotate=66, pos=0.32] {Dense Urban};
		
		\addplot[name path=f4, very thick, red, domain=0:90, smooth]{1/(1+27.1562*exp(-0.1228*(x-27.1562)))} node[below, rotate=63, pos=0.575] {Urban High-Rise};
		
	\end{axis}
}

\newcommand{\PlotSumRateFirst}[0]{

\begin{axis}[%
	xmin=-10,
	xmax=30,
	xlabel style={font=\color{white!15!black}},
	xlabel={SNR $\rho$ (dB)},
	ymin=0,
	ymax=45,
	ylabel style={font=\color{white!15!black}},
	ylabel={$R_{\Sigma}$ (bits/channel use)},
	axis background/.style={fill=white},
	legend style={at={(0.03,0.75), font=\small},anchor=west}, legend cell align={left}
	]

\addplot [color=blue, thick, mark=o, mark options={solid, blue}]
table[row sep=crcr]{%
-10	0.738897776899002\\
-5	1.94497431420996\\
0	4.47593786714395\\
5	8.66977922273794\\
10	14.2598068996894\\
15	20.6749625522245\\
20	27.4539672711771\\
25	34.365308179203\\
30	41.3206375470713\\
35	48.290108109054\\
40	55.2640741727842\\
45	62.2394642242723\\
50	69.2153048193606\\
55	76.1912879127774\\
60	83.1673160705163\\
65	90.1433584790843\\
70	97.1194053941842\\
};\addlegendentry{IA/ZF};

\addplot [color=red, thick, mark=asterisk, mark options={solid, red}]
table[row sep=crcr]{%
-10	0.602789327759192\\
-5	0.968195819454127\\
0	1.39145272277828\\
5	1.85022571249856\\
10	2.33157852163695\\
15	2.82379370254431\\
20	3.32007642320615\\
25	3.81772203740761\\
30	4.31580680096658\\
35	4.81403126901318\\
40	5.31229999936592\\
45	5.81058273509289\\
50	6.30886990054902\\
55	6.80715846689261\\
60	7.30544747624412\\
65	7.80373662568789\\
70	8.3020258194328\\
};\addlegendentry{BC};

\addplot [color=cyan, thick, mark=o, mark options={solid, cyan}]
table[row sep=crcr]{%
-10	0.576479776211711\\
-5	1.10503547252428\\
0	2.21923028168298\\
5	3.57309893101693\\
10	5.00387964681686\\
15	6.4759403637615\\
20	7.96331231654008\\
25	9.45578016377926\\
30	10.9498859427448\\
35	12.4445123636169\\
40	13.9393036953032\\
45	15.4341472033356\\
50	16.929007213731\\
55	18.423872442739\\
60	19.9187393221725\\
65	21.4136067224879\\
70	22.908474289621\\
};\addlegendentry{Blind BC};

\addplot [color=black, thick, mark=triangle, mark options={solid, rotate=270, black}]
table[row sep=crcr]{%
-10	1.29392433817612\\
-5	3.03124285602131\\
0	6.06469105071528\\
5	10.471111211561\\
10	16.0027911966242\\
15	22.1177138908406\\
20	28.465397017746\\
25	34.8991022808522\\
30	41.3625256483547\\
35	47.8357058037291\\
40	54.3120125702305\\
45	60.7893123822908\\
50	67.266926662053\\
55	73.7446404293502\\
60	80.2223856618081\\
65	86.7001408449534\\
70	93.1778991744865\\
}; \addlegendentry{P2P-TIN};

\addplot [color=green, thick, mark=triangle, mark options={solid, rotate=90, green}]
table[row sep=crcr]{%
-10	1.27776608751159\\
-5	3.02093116275315\\
0	6.11440353075532\\
5	10.7744277415566\\
10	16.6224354781353\\
15	23.0503325745317\\
20	29.6966287500478\\
25	36.4164315487457\\
30	43.1599634768097\\
35	49.9110490519757\\
40	56.6645283299912\\
45	63.4187650677777\\
50	70.1732413858689\\
55	76.9277934709555\\
60	83.6823695161749\\
65	90.436953138304\\
70	97.1915391564675\\
};\addlegendentry{2W};
\coordinate (spypoint) at (axis cs:22.75,33.5);
\coordinate (spyviewer) at (axis cs:10,33.5);
\spy[width=1.75cm,height=1.5cm] on (spypoint) in node [fill=white] at (spyviewer);				
\end{axis}}

\newcommand{\PlotSumRateSecond}[0]{
	
	\begin{axis}[%
		xmin=-10,
		xmax=30,
		xlabel style={font=\color{white!15!black}},
		xlabel={SNR $\rho$ (dB)},
		ymin=0,
		ymax=55,
		ylabel style={font=\color{white!15!black}},
		ylabel={$R_{\Sigma}$ (bits/channel use)},
		axis background/.style={fill=white},
		legend style={at={(0.03,0.75), font=\small},anchor=west}, legend cell align={left}
		]
		
		\addplot [color=blue, thick, mark=o, mark options={solid, blue}]
		table[row sep=crcr]{%
-10	0.848183697489422\\
-5	2.35154516641974\\
0	5.60491949608776\\
5	11.0400097543457\\
10	18.2782675703732\\
15	26.5571277694249\\
20	35.285638439721\\
25	44.176030842049\\
30	53.1200252987881\\
35	62.081229133977\\
40	71.0479015605935\\
45	80.0163059785173\\
50	88.9852583679082\\
55	97.9543840678849\\
60	106.923564576161\\
65	115.892762416611\\
70	124.861965738002\\
		};\addlegendentry{IA/ZF};
		
		\addplot [color=red, thick, mark=asterisk, mark options={solid, red}]
		table[row sep=crcr]{%
-10	1.37446660674027\\
-5	2.81679135100837\\
0	4.95390359938098\\
5	7.66191711474694\\
10	10.7128332038294\\
15	13.9828261899234\\
20	17.3833408970328\\
25	20.8410344501349\\
30	24.3191658247106\\
35	27.8040304018121\\
40	31.2910524600012\\
45	34.7787596398491\\
50	38.2666837621151\\
55	41.7546765164314\\
60	45.2426909769924\\
65	48.7307123019592\\
70	52.2187357976705\\
		};\addlegendentry{BC};
		
		\addplot [color=cyan, thick, mark=o, mark options={solid, cyan}]
		table[row sep=crcr]{%
-10	1.2506507433543\\
-5	2.65305535525071\\
0	4.92928984144183\\
5	7.86844919088694\\
10	11.2047232307964\\
15	14.7899706353127\\
20	18.5170771913563\\
25	22.3051667276909\\
30	26.114920100875\\
35	29.9317967194804\\
40	33.750954397088\\
45	37.5708362987113\\
50	41.3909475103825\\
55	45.2111312652738\\
60	49.0313379632966\\
65	52.8515519165217\\
70	56.6717681647725\\
		};\addlegendentry{Blind BC};
		
		\addplot [color=black, thick, mark=triangle, mark options={solid, rotate=270, black}]
		table[row sep=crcr]{%
-10	1.52422894379521\\
-5	3.4754944892428\\
0	6.62436160900862\\
5	10.8095624190234\\
10	15.6918539932721\\
15	20.0045602371833\\
20	25.0705165949082\\
25	30.4141367755434\\
30	35.8501333182302\\
35	41.3168134427354\\
40	46.7933862586407\\
45	52.2731079289795\\
50	57.7538274567404\\
55	63.2348627465953\\
60	68.7159979103922\\
65	74.1971646591909\\
70	79.6783413969223\\
		};\addlegendentry{P2P-TIN};
		
		\addplot [color=green, thick, mark=triangle, mark options={solid, rotate=90, green}]
		table[row sep=crcr]{%
-10	1.4924646707335\\
-5	3.45821980033852\\
0	6.94346148134581\\
5	11.9991958042854\\
10	18.1490432635367\\
15	24.9945377086339\\
20	32.2095466886752\\
25	39.5611552241775\\
30	46.9585131521164\\
35	54.3706185409424\\
40	61.787416392377\\
45	69.2057010503524\\
50	76.6244561707231\\
55	84.0433600936841\\
60	91.4623110750884\\
65	98.8812769379724\\
70	106.300247506823\\
		};\addlegendentry{2W};
\coordinate (spypoint) at (axis cs:8.5,16.35);
\coordinate (spyviewer) at (axis cs:-2.5,16.35);
\spy[width=1.75cm,height=1.5cm] on (spypoint) in node [fill=white] at (spyviewer);	
		
\end{axis}}

\newcommand{\PlotSumRateThird}[0]{
	
	\begin{axis}[%
		xmin=-10,
		xmax=30,
		xlabel style={font=\color{white!15!black}},
		xlabel={SNR $\rho$ (dB)},
		ymin=0,
		ymax=60,
		ylabel style={font=\color{white!15!black}},
		ylabel={$R_{\Sigma}$ (bits/channel use)},
		axis background/.style={fill=white},
		legend style={at={(0.03,0.75), font=\small},anchor=west}, legend cell align={left}
		]
		
		\addplot [color=blue, thick, mark=o, mark options={solid, blue}]
		table[row sep=crcr]{%
-10	0.867275378964067\\
-5	2.4238489553099\\
0	5.84585875966013\\
5	11.6384068828816\\
10	19.3995111272638\\
15	28.2898714148716\\
20	37.664983298711\\
25	47.2140204263374\\
30	56.8205729851344\\
35	66.4455833061756\\
40	76.0764580912305\\
45	85.7091901637914\\
50	95.3425098409986\\
55	104.976015363008\\
60	114.609579657094\\
65	124.243162536822\\
70	133.876751293874\\
		};\addlegendentry{IA/ZF};
		
		\addplot [color=red, thick, mark=asterisk, mark options={solid, red}]
		table[row sep=crcr]{%
-10	1.47722631895291\\
-5	3.14088285348674\\
0	5.69597362590896\\
5	9.04503831285198\\
10	12.8895019236951\\
15	17.0418201800227\\
20	21.3868406190712\\
25	25.8219348353545\\
30	30.2902515258121\\
35	34.7696378365593\\
40	39.2525843654718\\
45	43.7366628144904\\
50	48.2210998196576\\
55	52.7056502715058\\
60	57.1902366044752\\
65	61.6748342846645\\
70	66.1594355532213\\
		};\addlegendentry{BC};
		
		\addplot [color=cyan, thick, mark=o, mark options={solid, cyan}]
		table[row sep=crcr]{%
-10	1.43161715716843\\
-5	3.07497568706684\\
0	5.67073742428209\\
5	9.09097694593758\\
10	13.0274071790017\\
15	17.283586971958\\
20	21.7370565815964\\
25	26.2821486255763\\
30	30.8609632477519\\
35	35.4510066942176\\
40	40.0446608099306\\
45	44.6394628104379\\
50	49.2346284166437\\
55	53.8299090664269\\
60	58.4252261023393\\
65	63.0205546450557\\
70	67.6158868268378\\
		};\addlegendentry{Blind BC};
		
		\addplot [color=black, thick, mark=triangle, mark options={solid, rotate=270, black}]
		table[row sep=crcr]{%
-10	1.56064832764254\\
-5	3.54897644375668\\
0	6.68652835089478\\
5	10.7419929790247\\
10	15.3966762551212\\
15	19.3784548127956\\
20	23.9299594661848\\
25	28.8967692293493\\
30	33.9868924330924\\
35	39.1170749949068\\
40	44.2600951772568\\
45	49.4071941480068\\
50	54.5555849348057\\
55	59.7043844310033\\
60	64.8533131926898\\
65	70.0022828337142\\
70	75.151265401781\\
		};\addlegendentry{P2P-TIN};
		
		\addplot [color=green, thick, mark=triangle, mark options={solid, rotate=90, green}]
		table[row sep=crcr]{%
-10	1.52378166716398\\
-5	3.49711009639364\\
0	7.0088496604867\\
5	12.1128813672045\\
10	18.3470671467389\\
15	25.2736352036913\\
20	32.6480503376146\\
25	40.1990973678726\\
30	47.8106739872684\\
35	55.4419214317313\\
40	63.0794445850275\\
45	70.7189578921521\\
50	78.3591011036511\\
55	85.9994435647279\\
60	93.6398490396885\\
65	101.280274441952\\
70	108.920706145839\\
		};\addlegendentry{2W};
\coordinate (spypoint) at (axis cs:4,10);
\coordinate (spyviewer) at (axis cs:-3,21);
\spy[width=1.75cm,height=1.5cm] on (spypoint) in node [fill=white] at (spyviewer);

\end{axis}}

\begin{document}

\title{UAV-aided Multi-Way Communications}

\author{\IEEEauthorblockN{Jaber Kakar$^{*}$, Anas Chaaban$^{\dagger}$, Vuk Marojevic$^{\ddagger}$ and Aydin Sezgin$^{*}$}
	\IEEEauthorblockA{$^{*}$Institute of Digital Communication Systems,
		Ruhr-Universit{\"a}t Bochum, Germany\\$^{\dagger}$School of Engineering, University of British Columbia, Kelowna, Canada\\$^{\ddagger}$Bradley Dept. of Electrical and Computer Engineering, Virginia Tech, VA, USA\\
		Email: \{jaber.kakar, aydin.sezgin\}@rub.de, anas.chaaban@ubc.ca, maroje@vt.edu
	}}

\maketitle

\begin{abstract}
Multi-way and device-to-device (D2D) communications are currently considered for the design of future communication systems. Unmanned aerial vehicles (UAVs) can be effectively deployed to extend the communication range of D2D networks. To model the UAV-D2D interaction, we study a multi-antenna multi-way channel with two D2D users and an intermittently available UAV node. %The intermittency at the UAV node depicts the line-of-sight (LoS) dependency of air-to-ground channels for different ground environments at higher carrier frequencies. 
The performance in terms of sum-rate of various transmission schemes is compared. Numerical results show that for different ground environments, the scheme based on a combination of interference alignment, zero-forcing and erasure-channel treatment outperforms other schemes at low, medium and high SNRs and thus represents a viable transmission strategy for UAV-aided multi-way D2D networks.         
\end{abstract} 
\IEEEpeerreviewmaketitle

\section{Introduction}
\label{sec:intro}

Multi-way and device-to-device (D2D) %and full-duplex 
communications are important concepts in the design of next generation (5G) communication systems. Here, the term multi-way refers to multi-user networks where each node acts as a source, a destination and potentially a relay \cite{Chaaban15}. In D2D communications, users in close proximity are able to communicate with each other directly with minimal base-station (BS) or core network involvement \cite{Tehrani14}. Both multi-way and D2D communications can lead to significant improvement in the performance of communication networks. This includes, amongst others, an increase in spectral and energy efficiency as well as a decrease in delivery latency \cite{Kar17, Avestimehr10}.    

\begin{figure}[h]
	\vspace{0.5em}
	%\begin{turn}{90}
	\begin{tikzpicture}
	\SymModNew
	\end{tikzpicture}
	%\end{turn}
	%\vspace{-1em}
	\caption{\small A MIMO multi-way channel with the UAV as the intermittent node. $\boldsymbol{H}_{ij}$ is the channel matrix and $s_\ell\in\{0,1\}$ denotes the UAV availability at time instant $\ell$.}	
	\label{fig:SymMod}
\end{figure}
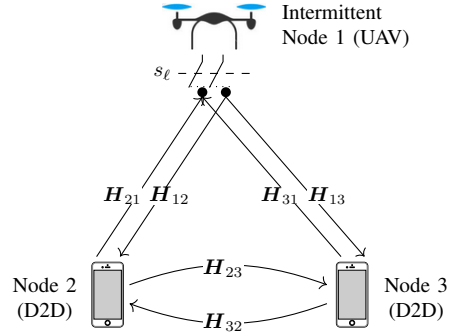

Further improvement in the performance of multi-way D2D networks for information dissemination is possible with the assistance of mobile base stations. These type of base stations may compensate for the limited communication range of D2D devices and provide the interface to the core network. In this context, unmanned aerial vehicles (UAVs) can be effectively used as mobile BSs to extend existing communication infrastructure \cite{Kakar15} and improve the quality of experience of users (e.g., providing UAV sensor data for safety in car-to-car communication). % For instance, Qualcomm recently completed its testing with a major US carrier of UAV communications through the 4G LTE cellular infrastructure \cite{Qualcomm17}. 
But the available bandwidths for envisioned UAV spectrum in the 5030-5091 MHz band \cite{Kerczewski13} may not suffice based on recent predictions on bandwidth requirements for UAVs \cite{Kakar17}. Spectrum at significantly higher frequencies (e.g. mmWaves \cite{Rupasinghe17}) are likely to be allocated, where communications heavily rely on line-
of-sight (LoS) links. \textcolor{black}{Due to this strong dependency on LoS, UAV communication links in this band will be intermittent, which necessitates studying the impact of intermittency on the communication performance. The implications of intermitent node availability on the system performance of interference and X-channels has been studied in prior works \cite{Karakus15, Yeh18}.}

In this work, we consider a UAV-assisted D2D setup where three nodes (a UAV and two D2D users) are equipped with multiple antennas and communicate with each other in a multi-way fashion (cf. Fig. \ref{fig:SymMod}). This setup represents a MIMO multi-way channel with three users %(or sometimes $\Delta$-channel) 
\cite{Chaaban16}. As opposed to most works on multi-way channels, we assume that the UAV might have intermittent connectivity, i.e., all incoming and outgoing links from and towards the UAV may be intermittent with some probability $\bar{\tau}$. This stems from the statistical nature of D2D$\leftrightarrow$UAV LoS availability, which is critical
for effective mmWave communication. \textcolor{black}{The (sum) degrees-of-freedom (DoF), i.e., the sum-capacity scaling over the signal-to-noise ratio (SNR) of the MIMO multi-way channel (Fig. \ref{fig:SymMod}) was studied in a previous work \cite{Chaaban17}}. %Interestingly, a scheme based on interference alignment (IA) and zero-forcing (ZF) was found to be sufficient for leveraging the DoF and is the basis of this work. 

However, the achievable sum-rate at low and medium SNR has not been studied, \textcolor{black}{and it remains to be seen which scheme performs best in practical SNR ranges.} Motivated by this, %and by the simplicity of the DoF-optimal non-adaptive scheme, 
this paper studies the achievable sum-rate at finite SNR. \textcolor{black}{We demonstrate that a scheme combining IA and ZF provides a good performance and even dominates various other schemes at even low SNR. Thus, this scheme is recommended for intermittent D2D$\leftrightarrow$UAV communication.}  

The rest of the paper is organized as follows. Section \ref{sec:sym_mod} describes the system model of the MIMO multi-way channel. A brief summary of schemes for the considered setup %, including the the previously proposed non-adaptive transmission scheme, 
is given in Section \ref{sec:non_adapt_scheme}. Section \ref{sec:uav_channel} establishes a relationship between intermittency and the statistical nature of air-to-ground channels. In Section \ref{sec:num_res}, numerical results are presented and discussed. Finally, Section \ref{sec:conclusion} provides concluding remarks.  
  
\textbf{Notation:} The notation $x_i^{n}$ denotes the concatenation $(x_{i,1},\ldots,x_{i,n})$ and $\boldsymbol{I}_N$ the $N\times N$ identity matrix. We use $\boldsymbol{X}\sim\mathcal{CN}(\boldsymbol{0},\boldsymbol{Q})$ to indicate that the random vector $\boldsymbol{X}$ follows a complex Gaussian distribution of zero mean and covariance matrix $\boldsymbol{Q}$. We write, respectively, $\lvert\lvert\boldsymbol{x}\rvert\rvert_2$, $\mathrm{N}(\boldsymbol{A})$ and $\textnormal{span}(\boldsymbol{A})$ to refer to the $\ell_2$ norm as well as the null and the column (sub)space of some matrix $\boldsymbol{A}$. Finally, we abbreviate $\max\{0,x\}$ by $x^{+}$.

\section{System Model}
\label{sec:sym_mod}

We consider a system illustrated in Fig. \ref{fig:SymMod} in which three MIMO \emph{full-duplex} nodes (two D2D nodes and a single UAV node) communicate with each other over a shared wireless channel. Each node $i\in\{1,2,3\}$ has a message $W_{ij}$ and $W_{ik}$ to be conveyed to the remaining two nodes $j$ and $k$ ($i\neq j\neq k$). Out of these three nodes, node $1$, which represents the UAV, is only intermittently present. We introduce the intermittency state parameter $s_\ell$ to denote if at time instant $\ell\in\{1,\ldots,n\}$ node $1$ is available ($s_\ell=1$) or unavailable ($s_\ell=0$). The intermittency state in our model is an independent and identically distributed (i.i.d.) random variable according to a Bernoulli distribution with probabilities $\mathbb{P}(s_\ell=1)=\tau$ and $\mathbb{P}(s_\ell=0)=\bar{\tau}\triangleq 1-\tau$. We denote the concatenation of states for $n$ consecutive channel uses by $s^{n}$. While $s^{n}$ is assumed to be known at node $1$, the state $s_\ell$ is only known causally at nodes $2$ and $3$.\footnote{This means that nodes $2$ and $3$ are not aware of $s_\ell$ prior to encoding the $\ell$-th transmission signal. However, through sensing their received signal they capture $s_\ell$ during reception.}

Each node $i\in\{1,2,3\}$ is equipped with $M_i$ transmit and receive antennas. The signal transmitted at time instant $\ell$ by the $i$-th node is represented by $\boldsymbol{x}_{i,\ell}\in\mathbb{C}^{M_i\times 1}$ which satisfies the average power constraint $\mathbb{E}[\lvert\lvert\boldsymbol{x}_{i,\ell}\rvert\rvert_{2}^{2}]\leq nP$. Using the same notation, the input-output relationship at all nodes are as follows\footnote{Note that $\boldsymbol{x}_{1,\ell}=\boldsymbol{0}$ if $s_\ell=0$.}
\begin{align}
	\boldsymbol{y}_{1,\ell}&=s_{\ell}\Big(\boldsymbol{H}_{21}\boldsymbol{x}_{2,\ell}+\boldsymbol{H}_{31}\boldsymbol{x}_{3,\ell}+\boldsymbol{z}_{1,\ell}\Big), \label{eq:rx_sig_node1} \\	
	\boldsymbol{y}_{2,\ell}&=\boldsymbol{H}_{12}\boldsymbol{x}_{1,\ell}+\boldsymbol{H}_{32}\boldsymbol{x}_{3,\ell}+\boldsymbol{z}_{2,\ell}, \label{eq:rx_sig_node2} \\	
	\boldsymbol{y}_{3,\ell}&=\boldsymbol{H}_{13}\boldsymbol{x}_{1,\ell}+\boldsymbol{H}_{23}\boldsymbol{x}_{2,\ell}+\boldsymbol{z}_{3,\ell}, \label{eq:rx_sig_node3} 
\end{align} where respectively, $\boldsymbol{H}_{ij}\in\mathbb{C}^{M_j\times M_i}$ is the channel (coefficient) matrix from node $j$ to node $i$ and $\boldsymbol{z}_{i,\ell}$ is a realization of the random Gaussian noise vector $\boldsymbol{Z}_{i,\ell}\sim\mathcal{CN}(\boldsymbol{0},\sigma^{2}\boldsymbol{I}_{M_i})$ being i.i.d. across time instants $\ell\in\{1,\ldots,n\}$. 

%Throughout this paper, the signal-to-noise ratio is abbreviated by SNR and defined by $\rho=P/\sigma^{2}$. 
The SNR is defined as $\rho=P/\sigma^{2}$. For the system under study, we assume that $M_1\geq M_2\geq M_3$.\footnote{Other antenna configuration of the intermittent channel setup have been studied in \cite{Neu17}.} All nodes are assumed to have perfect channel state information (CSI). Encoding, decoding and the error probability are defined in the standard Shannon sense \cite{Cover06} and its description is omitted for the sake of brevity. We indicate the achievable rates with respect to messages $W_{ij}$ by $R_{ij}(\rho)$ and the corresponding \emph{sum-capacity} by $C_{\Sigma}(\rho)$. The (sum) DoF measures the slope in the sum-capacity through comparison with the (approximate) capacity of a point-to-point channel ($\approx \log_2(\rho)$) at sufficiently high SNR. It is given by \cite{KakarEnt17}
\begin{equation}
	\label{eq:sum_DoF}
	d_{\Sigma}=\lim_{\rho\rightarrow\infty}\frac{C_{\Sigma}(\rho)}{\log_2(\rho)}.
\end{equation} In the next section, we will present multiple \emph{non-adaptive} schemes, including the DoF-optimal scheme based on IA and ZF, for the considered multi-way setup. Non-adaptive (or restricted) in the case of the multi-way channel means that the encoder of the $i$-th node for time instant $\ell$ constructs $\boldsymbol{x}_{i,\ell}$ without relying on past channel outputs $\boldsymbol{y}_{i}^{\ell-1}$, or in short
\begin{equation}
	\label{eq:non_adapt_enc}
	\begin{cases}
		\boldsymbol{x}_{1,\ell}&=\epsilon_{1,\ell}(W_{12},W_{13},s^{n}) \\
		\boldsymbol{x}_{j,\ell}&=\epsilon_{j,\ell}(W_{ji},W_{jk})
	\end{cases}
\end{equation} for $j\in\{2,3\},i\neq j\neq k$, where $\epsilon_{i,\ell}(\cdot)$ is the encoding function of node $i$ at time $\ell$.

\section{Schemes under Investigation}
\label{sec:non_adapt_scheme}

We present multiple schemes, including a scheme based on interference alignment and zero-forcing in sub-section \ref{subsec:ZF_IA} as well time-sharing-based schemes in sub-section \ref{subsec:time_sharing}. 

\subsection{IA/ZF Scheme}
\label{subsec:ZF_IA}
 
Each node splits its message into distinct sub-messages where a sub-message is either sent by zero-forcing (ZF) or interference alignment (IA). Then it uses superposition coding to encode the sub-messages. Specifically, the transmit signal at the $i$-th node becomes
\begin{equation}
	\boldsymbol{x}_{i,\ell}=\sum_{j=1,j\neq i}^{3}\Big(\boldsymbol{V}_{ij}^{\mathsf{IA}}\boldsymbol{x}_{ij,\ell}^{\mathsf{IA}}+\boldsymbol{V}_{ij}^{\mathsf{ZF}}\boldsymbol{x}_{ij,\ell}^{\mathsf{ZF}}\Big), 
\end{equation} where $\boldsymbol{V}_{ij}^{q}\in\mathbb{C}^{M_i\times a_{ij}^{q}},q\in\{\mathsf{ZF},\mathsf{IA}\}$, represents the transmit beamforming matrix and $a_{ij}^{q}$ the number of independent streams to be allocated for the ZF or IA sub-message of $W_{ij}$. Each beamforming matrix is independently chosen to avoid overlaps in the transmit signal space. Specifically, the ZF beamforming matrix is chosen such that $\boldsymbol{V}_{ij}^{\mathsf{ZF}}\in\mathrm{N}(\boldsymbol{H}_{ik})$ for distinct $i,j,k\in\{1,2,3\}$. ZF is possible as long as $a_{ij}^{\mathsf{ZF}}$ does not exceed the null space dimension, i.e., 
\begin{equation}
	a_{ij}^{\mathsf{ZF}}\leq(M_i-M_k)^{+}. 
\end{equation} Node $i$ receives a noise-distorted linear observation of the desired signals (denoted by superscript $(d)$) $\boldsymbol{x}_{i,\ell}^{(d),q}=[{\boldsymbol{x}_{ji,\ell}^{q}}^{T},{\boldsymbol{x}_{ki,\ell}^{q}}^{T}]^{T},q\in\{\mathsf{ZF},\mathsf{IA}\}$ and undesired signals (denoted by $(\bar{d})$) $\boldsymbol{x}_{i,\ell}^{(\bar{d})}=[{\boldsymbol{x}_{jk,\ell}^{\mathsf{IA}}}^{T},{\boldsymbol{x}_{kj,\ell}^{\mathsf{IA}}}^{T}]^{T}$, or in short,  
\begin{equation}
	\boldsymbol{y}_{i,\ell}=\boldsymbol{H}_i^{(d)}
	\begin{bmatrix}
		\boldsymbol{x}_{i,\ell}^{(d),\mathsf{ZF}} \\ \boldsymbol{x}_{i,\ell}^{(d),\mathsf{IA}} 
	\end{bmatrix}
	+\boldsymbol{H}_i^{(\bar{d})}\boldsymbol{x}_{i,\ell}^{(\bar{d})}+\boldsymbol{z}_{i,\ell},
\end{equation} with effective channel matrices
\begin{align}
	\boldsymbol{H}_i^{(d)}
		&=[\boldsymbol{H}_{ji}\boldsymbol{V}_{ji}^{\mathsf{ZF}},\boldsymbol{H}_{ki}\boldsymbol{V}_{ki}^{\mathsf{ZF}},\boldsymbol{H}_{ji}\boldsymbol{V}_{ji}^{\mathsf{IA}}, \boldsymbol{H}_{ki}\boldsymbol{V}_{ki}^{\mathsf{IA}}],\\
	\boldsymbol{H}_i^{(\bar{d})}
		&=[\boldsymbol{H}_{ji}\boldsymbol{V}_{jk}^{\mathsf{IA}}, \boldsymbol{H}_{ki}\boldsymbol{V}_{kj}^{\mathsf{IA}}].	
\end{align} At the $i$-th node, post-coding matrices $\boldsymbol{T}_{ji}^{q}$ are applied to remove the contribution of all components except of the desired component $\boldsymbol{x}_{ji,\ell}^{q}$. This requires that the columns of the concatenated matrices $\boldsymbol{H}_i^{(d)}$ and $\boldsymbol{H}_i^{(\bar{d})}$ are linearly independent. Nevertheless, the goal is to reserve most of the received signal space for the desired components, i.e., minimize the effect of the interference subspace $\textnormal{span}\big(\boldsymbol{H}_i^{(\bar{d})}\big)$ by maximizing the dimension $\bar{a}_{jk}^{\mathsf{IA}}$ of the intersection subspace 
\begin{equation}
	\label{eq:intersection_subspace}
	\textnormal{span}\Big(\boldsymbol{H}_{ji}\boldsymbol{V}_{jk}^{\mathsf{IA}}\Big)\cap 	\textnormal{span}\Big(\boldsymbol{H}_{ki}\boldsymbol{V}_{kj}^{\mathsf{IA}}\Big). 
\end{equation} This can be achieved by appropriately designing $\boldsymbol{V}_{jk}^{\mathsf{IA}}$ and $\boldsymbol{V}_{kj}^{\mathsf{IA}}$. However, it is easy to see that the dimension of this intersection subspace cannot be arbitrarily increased. In fact, it is upper bounded according to
\begin{equation}
	\bar{a}_{jk}^{\mathsf{IA}}\leq \min\{a_{jk}^{\mathsf{IA}},a_{kj}^{\mathsf{IA}},(M_j+M_k-M_i)^{+}\}.
\end{equation} When it comes to the decoding of signals $\boldsymbol{x}_{1j,\ell}^{q}$ or $\boldsymbol{x}_{j1,\ell}^{q}$, i.e., signals sent or received by node $1$, the achievable rates scale with the probability $\tau$ as these signals experience an erasure channel \cite{Cover06}. Formulating an optimization problem with respect to $a_{ij}^{q}$ and $\bar{a}_{ij}^{\mathsf{IA}}$, one can show that
\begin{equation}
	\label{eq:opt_DoF_3WC}
	d_{\Sigma}=2\tau M_2+2\bar{\tau}M_3.
\end{equation} 

\subsection{Time-Sharing-Based Schemes}
\label{subsec:time_sharing}

\begin{figure}[h]
	\centering
	\begin{tikzpicture}[scale=0.85]
	\PlotLoS
	\end{tikzpicture}
	\caption{$\mathbb{P}(\text{LoS})$ for suburban, urban, dense urban and urban high-rise ground environments as a function of the elevation angle $\theta\in[0,90]$.}
	\label{fig:Los_prob}
\end{figure}
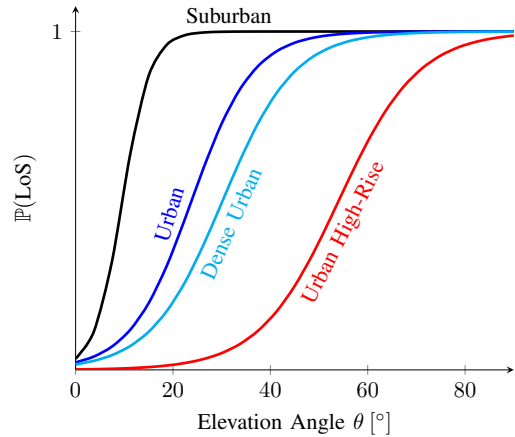  

The time-sharing based schemes, we consider are: 
\begin{enumerate}[wide, labelwidth=!, labelindent=0pt]
	\item \textbf{BC:} This scheme uses MIMO broadcasting (BC) in combination with time sharing, i.e., only in time instants where $s_\ell=1$, node $i$ is active (in $\nicefrac{1}{3}$ fractions) and broadcasts its messages to the remaining two nodes. Exploiting the duality between MIMO-BC and MIMO-MAC, the optimal transmit covariance matrix for the broadcast transmissions is determined based on the iterative water-filling algorithm of \cite{Jindal02}. 
	\item \textbf{"Blind" BC:} Node $i,i\in\{1,2,3\}$ transmits for $\nicefrac{1}{3}$ time fractions. Naively each node $i$, assumes $s_\ell=1$ and broadcasts the messages to the remaining users (even though it might be that $s_\ell=0$). The covariance matrix is optimized by iterative water-filling. 
	\item \textbf{P2P-TIN:} This scheme uses TDMA in connection with treating interference as noise (TIN) and water-filling for single-user point-to-point (P2P) channels \cite{Telatar99}. Specifically, in all time instants where $s_\ell=1$ the nodes communicate with each other in the following two modes:
	\begin{itemize}
		\item Node $1\rightarrow$ Node $2\rightarrow$ Node $3\rightarrow$ Node $1$ 
		\item Node $1\rightarrow$ Node $3\rightarrow$ Node $2\rightarrow$ Node $1$ 
	\end{itemize} Each communication mode occurs half the time when $s_\ell=1$. When $s_\ell=0$, on the other hand, only nodes 2 and 3 communicate in a two-way fashion. 
	%\item \textbf{BS IV:} This schemes extends BS III by applying water-filling on top of it.   
	\item \textbf{2W:} In this scheme, only two nodes are active at a time and communicate in an optimized two-way (2W) fashion. Specifically, when $s_\ell=0$, nodes 2 and 3 communicate whereas when $s_\ell=1$, the communication of the three possible node pairs happens each one third of the remaining time.    
\end{enumerate} 
Note that with the exception of blind BC, all other time-sharing-based schemes explicitly assume the knowledge of the state sequence $s^{n}$ at \emph{all} nodes. In contrast, the proposed IA/ZF scheme does not need knowledge of $s_\ell$ at nodes $j\in\{2,3\}$ when transmitting $\boldsymbol{x}_{j,\ell}$. 

In the next section, we establish a link between intermittency probabilities $\tau$ and $\bar{\tau}$ of the multi-way channel with LoS and non LoS (NLoS) probabilities of state-of-the-art UAV channel models.
\begin{figure*}[h]
	%\begin{figure}[h]
	\hspace{-1cm}
	\centering
	\begin{minipage}[c]{0.42\linewidth}
		\begin{subfigure}[c]{0.75\textwidth}
			\centering
			\begin{tikzpicture}[scale=0.95, spy using outlines={circle, magnification=3,connect spies}]
			\PlotSumRateFirst
			\end{tikzpicture}
			\caption{$\tau=0.1$}
			\label{fig:sum_rate_UHR}
		\end{subfigure}
	\end{minipage}
	\hspace{1cm}
	%\hfill
	\begin{minipage}[c]{0.42\linewidth}
		\begin{subfigure}[c]{0.75\textwidth}
			\centering
			\begin{tikzpicture}[scale=0.95, spy using outlines={circle, magnification=3,connect spies}]
			\PlotSumRateSecond
			\end{tikzpicture}
			\caption{$\tau=0.7$}
			\label{fig:sum_rate_DU}
		\end{subfigure}
	\end{minipage}
	\vfill
	\begin{minipage}[c]{0.45\linewidth}
		\begin{subfigure}[c]{0.80\textwidth}
			\centering
			\begin{tikzpicture}[scale=0.95, spy using outlines={circle, magnification=3,connect spies}]
			\PlotSumRateThird
			\end{tikzpicture}
			\caption{$\tau=0.9$}
			\label{fig:sum_rate_U}
		\end{subfigure}
	\end{minipage}
	\caption{Sum-rate of the node-intermittent multi-way channel for $\tau\in\{0.1,0.7,0.9\}$ at fixed elevation angle $\theta\approx 37.5^{\circ}$ for (a) urban high-rise ($\tau^{\mathsf{UHR}}=0.1$), (b) dense-urban ($\tau^{\mathsf{DU}}=0.7$) and (c) urban ($\tau^{\mathsf{U}}=0.9$) ground environments. Recall that the BC, P2P-TIN, and 2W schemes operate with apriori knowledge of $s^n$ at all nodes, contrary to the Blind BC and IA/ZF schemes which do not.}
	\label{fig:sum_rate}
\end{figure*}
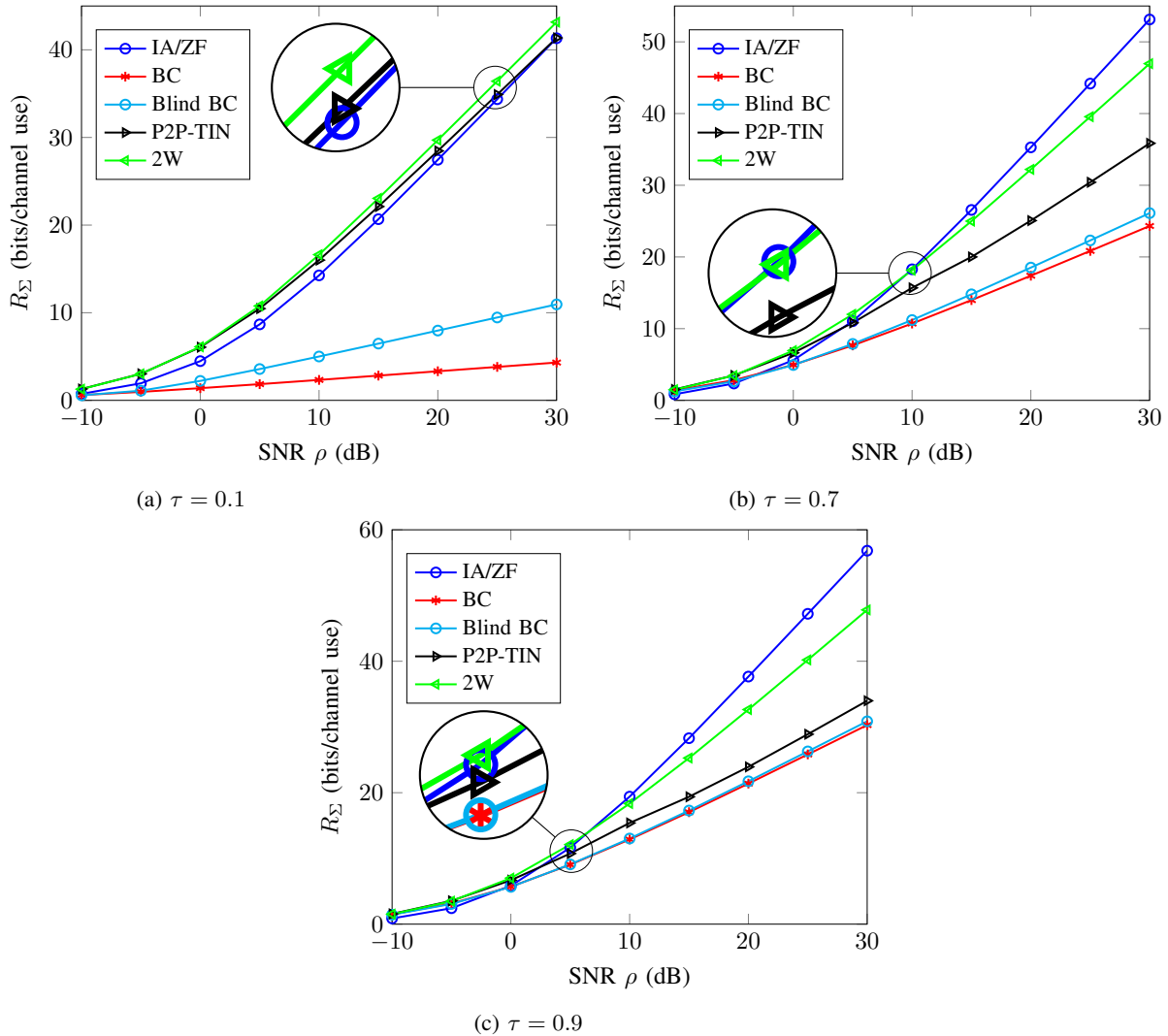 

\section{UAV Communication: LoS and NLoS}  
\label{sec:uav_channel}

In the existing literature of UAV channel models\footnote{Here, we mainly refer to path loss models for low and high altitude platforms.}, the large-scale effect of the channel in terms of path loss is mainly governed by the UAV and the ground device (e.g. D2D user) having LoS or NLoS links. To this end, many works in the existing literature deploy a probabilistic path loss model (e.g. \cite{Feng06,Holis08,Hourani14}). In these models, the path loss of LoS connections is described by the free space path loss whereas NLoS connections suffer from an additional additive excessive path loss $\eta$ that may depend on the carrier frequency $f$ and the elevation angle $\theta$ (in $^{\circ}$) \cite{Holis08,Hourani14}. Clearly, %$\theta=\nicefrac{180}{\pi}\sin^{-1}\big(\frac{h}{d}\big)$ 
\begin{equation}
	\label{eq:theta}
	\theta=\frac{180}{\pi}\sin^{-1}\bigg(\frac{h}{d}\bigg),
\end{equation} 
with $h$ being the UAVs altitude and $d$ the UAV-to-ground-user distance. In this context, LoS and NLoS connections occur with probabilities \cite{Hourani14}
\begin{align}
	\mathbb{P}(\text{LoS})&=\frac{1}{1+a_{1}e^{-a_2(\theta-a_1)}},\label{eq:plos}\\
	\mathbb{P}(\text{NLoS})&=1-\mathbb{P}(\text{LoS}),\label{eq:pnlos}
\end{align} 
%\begin{align}
%\mathbb{P}(\text{LoS})&=\frac{1}{1+a_{1}e^{-a_2(\theta-a_1)}}
%\end{align} and $\mathbb{P}(\text{NLoS})=1-\mathbb{P}(\text{LoS})$, 
where the constants $a_1$ and $a_2$ depend on the ground environment (rural, urban, dense urban, urban high-rise)\footnote{ITU-R defines typical realizations of these environments based on three parameters: i) average normalizing building area, ii) mean number of buildings per unit area and iii) the Rayleigh scale parameter for building heights \cite{ITU-R12}.}. Fig. \ref{fig:Los_prob} shows the LoS probability of these four environments as a function of $\theta$. It is intuitive, that the LoS probability increases with increasing $\theta$ due to the reduced influence of signal blockage. We approximate the probabilities $\mathbb{P}(\text{LoS})$ and $\mathbb{P}(\text{NLoS})$\footnote{This is mainly due to the fact there is an excessive path loss $\eta$.} as follows
%as $\mathbb{P}(\text{LoS})\approx\tau$ and $\mathbb{P}(\text{NLoS})\approx\bar{\tau}$\footnote{This is mainly due to the fact there is an excessive path loss $\eta$.} for instance, note that
\begin{equation*}
 \mathbb{P}(\text{LoS})\approx\tau,\:\mathbb{P}(\text{NLoS})\approx\bar{\tau}. 
\end{equation*} 
For instance, at $\theta=37.5^{\circ}$ in \underline{u}rban (U), \underline{d}ense \underline{u}rban (DU) and \underline{u}rban \underline{h}igh-\underline{r}ise (UHR) environments, the intermittency probability can vary significantly from $\tau^{\mathsf{U}}\approx 0.9$, $\tau^{\mathsf{DU}}\approx 0.7$ to $\tau^{\mathsf{UHR}}\approx 0.1$ (Fig. \ref{fig:Los_prob}). 

Next, we use these intermittency probabilities to compare the IA/ZF scheme with the time-sharing schemes. % in terms of the achievable sum-rate of the multi-way channel for different ground environments.   

\section{Numerical Results}
\label{sec:num_res}
We compare the schemes of section \ref{sec:non_adapt_scheme} in terms of the sum-rate (in bits per channel use) versus SNR for $M_1=5,M_2=3$ and $M_3=2$. 

The achievable sum-rate of all schemes is shown in Fig. \ref{fig:sum_rate} for $\tau=0.1$ (Fig. \ref{fig:sum_rate_UHR}), $\tau=0.7$ (Fig. \ref{fig:sum_rate_DU}) and $\tau=0.9$ (Fig. \ref{fig:sum_rate_U}). From the plot of Fig. \ref{fig:sum_rate}, it can be seen that the IA/ZF scheme outperforms various time-sharing schemes (BC, blind BC and P2P-TIN) already at medium SNR ($\rho\gtrapprox 15$ dB). This is mainly so because the slope of the achievable sum-rate is close to the optimal slope ($6\tau+4\bar{\tau}$, cf. Eq. \eqref{eq:opt_DoF_3WC}) at SNR values exceeding $15$ dB. While BC and blind BC perform poorly, irrespective of $\tau$, the performance of P2P-TIN improves with decreasing $\tau$ (cf. Fig. \ref{fig:sum_rate_UHR} vs. \ref{fig:sum_rate_U}). The reason for this is that as $\tau$ decreases, nodes 2 and 3 communicate more frequently in an optimized two-way fashion. The achievable sum-rate of 2W increases with $\tau$ because the two-way channel between nodes $1$ and $2$ attains a higher multiplexing gain ($\min\{M_1,M_2\}\geq M_3$) than any other two-way channel pair. With increasing $\tau$, these two nodes are able to communicate with higher frequency and as such increase the overall achievable sum-rate. For $\tau=0.1$, we can see that schemes P2P-TIN and 2W perform similarly or even slightly better than the IA/ZF scheme. For small $\tau$, node $1$ is frequently unavailable; consequently, the benefits of IA are seen at significantly high SNR at which the IA/ZF scheme outperforms 2W. 

We recall that in both P2P-TIN and 2W, the knowledge of $s_\ell$ is assumed (at nodes 2 and 3) prior to the transmission at time instant $\ell$. This is \emph{not} the case for the IA/ZF scheme. This means that the non-adaptive IA/ZF scheme is suitable for all ground environments because being robust and agnostic to the availability of the UAV.  

\section{Conclusion}
\label{sec:conclusion}

In this paper, we investigate the performance of UAV-aided multi-way networks. To this end, we analyze the achievable sum-rate of a node-intermittent multi-way channel consisting of two users and an intermittently available UAV communication node. %The intermittency at the UAV is due to the line-of-sight dependency of air-to-ground channels. 
To account for the intermittently available UAV node, we deploy a non-adaptive scheme based on interference alignment, zero-forcing and erasure-channel treatment. Numerical results show that this scheme performs significantly better than various other schemes for UAV LoS availability levels ranging from suburban to dense urban ground environments. Considering intermittent availability of aerial nodes, allows scaling the system to multiple D2D users and UAVs.

\bibliographystyle{IEEEtran}
\bibliography{content/bibliography}
\balance
 
\end{document}